\documentclass[sigconf]{acmart}
\copyrightyear{2026}
\acmYear{2026}
\setcopyright{cc}

\acmConference[CIKM '26]
{Proceedings of the 35th ACM International Conference on Information and Knowledge Management}
{November 7--11, 2026}
{Rome, Italy.}

\acmBooktitle{Proceedings of the 35th ACM International Conference on Information and Knowledge Management (CIKM '26), November 7--11, 2026, Rome, Italy}

\acmISBN{979-8-4007-2539-5/2026/11}
\setcctype{by}
\acmDOI{10.1145/3799682.3840215}
\newcommand{\best}[1]{\textbf{#1}}
\newcommand{\ndcg}{nDCG@10}
\newcommand{\mapk}{MAP@1000}
\newcommand{\rk}{Recall@1000}

\newcommand{\secondbest}[1]{\underline{#1}}
\newcommand{\sys}{\textsc{QueryRoute}\xspace}
\usepackage{xspace}
\newcommand{\primarybackbone}{\texttt{Qwen2.5-7B-Instruct}\xspace}
 
\usepackage{algorithm}
\usepackage{algpseudocode}
\usepackage{amsmath}
\usepackage{amsfonts}
\usepackage{booktabs}
\usepackage{balance}
\usepackage{microtype}
\usepackage{amsmath}
\usepackage{graphicx}
\usepackage{pgfplots}
\usepackage[utf8]{inputenc}  % Most likely already there
\usepackage[T1]{fontenc}
\usepackage{textcomp}
\usepackage{enumitem}
\usepackage{multirow}
\usepackage{makecell}
\usepackage[most]{tcolorbox}
\usepackage{xcolor}

\usepackage{algorithm}
\usepackage{algpseudocode}
\usepackage{amsmath}      % for \begin{cases}, \dfrac, \mathbb

\usepackage{amssymb}      % for \mathbb{I} (indicator)

\begin{document}

%%
% The "title" command has an optional parameter,
\title{Route Me If You Can:\\A Benchmark for Query Reformulation Selection}

% \affiliation{%
%   \institution{University of Toronto}
% }

% \affiliation{%

% \affiliation{%

% \affiliation{%

% \affiliation{%

% \affiliation{%

% \affiliation{%
\settopmatter{authorsperrow=4}
\author{Hai Son Le}
% \email{haison.le@torontomu.ca}
\orcid{0009-0003-2240-0451}
\affiliation{%
  \institution{Toronto Metropolitan University}
  \city{Toronto}
  \country{Canada}
  }

\author{Negar Arabzadeh}
\orcid{0000-0002-4411-7089}
\affiliation{%
  \institution{University of California, Berkeley}
  \city{Berkeley}
  \country{USA}
  }

\author{Amin Bigdeli}
% \email{aminbigdeli97@gmail.com}
\orcid{0009-0003-8977-9312}
\affiliation{%
  \institution{University of Waterloo}
  \city{Waterloo}
  \country{Canada}
  }

\author{Radin Hamidi Rad}
\orcid{0000-0002-9044-3723}
\affiliation{%
  \institution{Mila - Quebec AI Institute}
  \city{Montreal}
  \country{Canada}
  }

\author{Sajad Ebrahimi}
\orcid{0009-0003-1630-3938}
\affiliation{%
  \institution{University of Toronto}
\city{Toronto}
  \country{Canada}
  }

\author{Charles L. A. Clarke}
\orcid{0000-0001-8178-9194}
\affiliation{%
  \institution{University of Waterloo}
    \city{Waterloo}
  \country{Canada}
  }

\author{Ebrahim Bagheri}
% \email{ebrahim.bagheri@utoronto.ca}
\orcid{0000-0002-5148-6237}
\affiliation{%
 \institution{University of Toronto}
 \city{Toronto}
 \country{Canada}
 }



% %%
% %% By default, the full list of authors will be used in the page
% %% headers. Often, this list is too long, and will overlap
% %% other information printed in the page headers. This command allows
% %% the author to define a more concise list
% %% of authors' names for this purpose.
\renewcommand{\shortauthors}{Hai Son Le et al.}


\begin{abstract}
LLM-based query reformulation can improve retrieval, but no single reformulation strategy is consistently optimal across queries, domains, retrievers, or model backbones. This creates an inference-time decision problem: \textit{``Given an original query and a pool of candidate reformulations, which one should be issued to the retriever?''}. Existing studies are hard to compare because they use different reformulator pools, retrievers, relevance signals, training labels, and evaluation metrics. We introduce \sys, a benchmark that freezes the expensive artifacts needed to study this decision reproducibly: original queries, generated variants, ranked lists under multiple retrievers, retrieval scores, and per-query oracle labels. The benchmark contains 3,757 queries, 11 candidate systems, five reformulator backbones, and three retrievers across TREC DL, BEIR, and BRIGHT, yielding 619,905 retrieval outcomes. We benchmark supervised classification, routing, QPP, and LLM-as-judge selectors. Results show substantial oracle headroom over fixed reformulators, but current selectors recover only part of it; selector rankings change across retrievers, and similar mean effectiveness can hide different query-level behavior. The released artifacts and evaluation harness allow future selectors to be compared without regenerating variants, rerunning retrieval, or rebuilding judge pipelines. Code and data are available at \url{https://github.com/haisonle001/QueryRoute}.

\end{abstract}
% \pagestyle{plain}
% # will add number later




\maketitle

\begin{figure}[t]
  \centering
  \includegraphics[width=0.74\columnwidth,height=0.25\textheight]{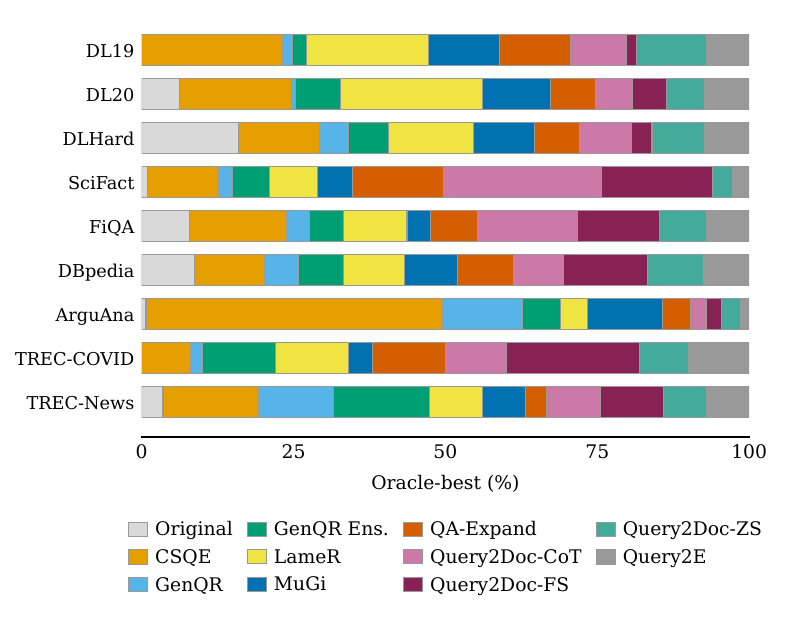}
  \vspace{-1em}
\caption{Oracle-winner distribution across different datasets with the \primarybackbone{} generator. Different reformulators win on different datasets, highlighting the opportunity for query-reformulation routing.}
  \vspace{-2em}
  \label{fig:oracle-winners}
\end{figure}

% \vspace{-1em}
\section{Introduction}

Query reformulation has long been a central component of information retrieval. Classical IR has studied manual, automatic, and interactive query expansion, relevance feedback, pseudo-relevance feedback, term selection, and query refinement as ways to bridge vocabulary mismatch and improve retrieval effectiveness~\cite{rocchio1971relevance,bhogal2007review,qiu1993concept}. Selecting among query variants has been studied for decades under headings such as query variation and query-by-query selection~\cite{uqv100, Benham_2019,cronen_selective}. LLM-based reformulation revives this long-standing problem in a new form. Modern pipelines can generate many variants of the same information need using different prompt templates, reformulation objectives, decoding strategies, and model backbones. These variants may include keyword expansions, pseudo-documents, answer-style passages, question decompositions, or corpus-grounded rewrites~\cite{Genqr,Genqrensemble,query2doc,qa-expand,mugi,lamer,csqe}. Recent LLM-based reformulation studies show that no single strategy is consistently optimal across queries, datasets, retrievers, or model backbones~\cite{querygym-repro}. Thus, the question is not only how to generate better reformulations, but also how to choose among available variants for a particular query and retrieval setting.

% We call this problem \textbf{query variant selection}. 
We revisit this selection problem under LLM-based reformulation and study it as query reformulation selection. Given an original query and a fixed pool of candidate variants, a selector must choose one to issue to a retriever, without test-set relevance labels. The selected query determines the ranked list returned or passed downstream. This is distinct from evaluating reformulators by average effectiveness: a fixed reformulator can be strong on average while failing on many queries, whereas a selector can improve retrieval by choosing different reformulators for different queries. Figure~\ref{fig:oracle-winners} illustrates this: oracle-best reformulators are distributed across methods and datasets. Query reformulation selection is therefore an intra-query decision problem: all candidates express the same need but induce different outcomes. Several recent works point toward this view. QPP-based variant selection treats query performance prediction as a mechanism for choosing among variants~\cite{arabzadeh2026qpp}, and learned routing methods for selecting among LLMs or systems can be adapted to reformulators~\cite{routellm,hybrid_llm,best_route}. These methods show selection matters but are hard to compare, since each builds its own candidate pool, retriever configuration, labels, and protocol. The missing resource is therefore not another reformulator or selector, but a benchmark that freezes the expensive intermediate artifacts needed to study selection.

We present \sys{}, a resource and benchmark for studying this revived selection problem reproducibly. \sys{} freezes the expensive artifacts needed for comparison: original queries, LLM-generated variants, ranked lists under multiple retrievers, retrieval scores, and per-query oracle labels. The benchmark covers 3,757 queries across TREC DL, BEIR, and BRIGHT-style settings; 10 LLM-based reformulators plus the original query; 5 reformulator backbones; and 3 retrievers, yielding 619,905 query--variant--retriever outcomes.  Our empirical characterization shows substantial oracle headroom over fixed reformulation strategies, confirming that there is room for improved per-query selection. At the same time, current selectors recover only part of this potential: selector behavior depends strongly on domain and retriever choice, and mean effectiveness can hide substantially different routing behavior. \sys{} enables future query-reformulation selectors to be evaluated without regenerating variants, rerunning retrieval, or rebuilding judge pipelines.
The contributions of this resource are:
% We introduce \sys, a resource and benchmark for reproducible query variant selection. For each query, \sys{} provides the original query, a fixed set of generated variants, ranked lists under multiple retrievers, retrieval scores, qrels-derived oracle labels. A selector reads the artifacts and outputs one candidate identifier per query. The evaluation harness joins these predictions with the frozen ranked lists and reports both retrieval effectiveness and decision-quality metrics. This design makes it possible to evaluate a new selector without rerunning generation, retrieval, or judging.

% \sys{} spans three retrieval regimes. First, TREC DL and DL-Hard represent high-resource web search with graded relevance judgments. Second, BEIR datasets test heterogeneous zero-shot retrieval across scientific, argumentative, biomedical, financial, encyclopedic, and news domains. Third, BRIGHT-style reasoning-intensive subsets can be included as an optional stress-test regime, where retrieval often requires multi-step reasoning rather than surface lexical or semantic matching. Across these regimes, \sys{} varies reformulator method, LLM backbone, and retriever, enabling controlled analysis of whether selector behavior transfers across domains and retrieval paradigms. 

\begin{itemize}[leftmargin=1.5em, itemsep=0pt, topsep=0pt]
    \item We revisit query variant selection in the context of LLM-based reformulation and treat it as a standardized finite-action benchmark over fixed candidate pools, retrievers, target metrics, and evaluation splits.
    % \item We release a frozen multi-regime artifact suite containing candidate variants, ranked lists, retrieval scores, oracle labels, selector splits, QPP features, and optional LLM-judge labels.
    \item We release a frozen benchmark covering 3,757 queries, 11 candidate systems per query, 5 LLM backbones, and 3 retrievers, for a total of 619,905 retrieval outcomes.
    \item We provide a standardized evaluation harness that reports both retrieval effectiveness and decision quality, including selection accuracy, near-oracle rate, regret, oracle-gap closure, help/hurt rate, pairwise accuracy, rank correlation, and selection entropy.
    \item We benchmark selectors from supervised classification, learned routing, QPP, and LLM-as-judge families, showing oracle headroom that no fixed reformulator closes.
\end{itemize}

\vspace{-1em}
\section{Related Work}
% \noindent \textbf{LLM-based query reformulation.}
% LLM-based query reformulation methods use generative models to rewrite, expand, or enrich an input query before retrieval. Keyword-level methods prompt the model to generate high-impact expansion terms or semantically related phrases, as in GenQR, GenQR-Ensemble, and Query2E~\cite{Genqr,Genqrensemble,exploringresearch}. Document-level methods synthesize answer-style pseudo-documents or longer contextual passages, as in Query2Doc, QA-Expand, and MuGI~\cite{query2doc,qa-expand,mugi}. Corpus-grounded methods condition generation on initially retrieved evidence or collection-level signals, as in LameR and CSQE~\cite{lamer,csqe}. These methods often improve aggregate retrieval effectiveness, especially under lexical retrieval, but recent reproducibility work shows that gains are sensitive to the retriever, dataset, model backbone, and generation configuration~\cite{querygym-repro}.

\noindent\textbf{LLM-based query reformulation.} LLM reformulators rewrite or enrich queries through keyword generation, pseudo-document generation, multi-question expansion, and corpus-grounded rewriting~\cite{Genqr,Genqrensemble,query2doc,qa-expand,mugi,lamer,csqe}. These methods often improve average retrieval effectiveness, especially for sparse retrieval, but gains vary across queries, datasets, retrievers, and model backbones~\cite{querygym-repro}. This variability motivates per-query selection rather than a single fixed reformulator.

\noindent \textbf{Inference-time selection.} A separate line of work decides which reformulation to use at inference time: AdaRewriter ranks rewrites with a trained reward model~\cite{adarewriter}, QueStER learns a query-specification policy for a fixed lexical retriever~\cite{satouf2025quester}, and ReFormeR applies explicit reformulation patterns via a lightweight per-query selector~\cite{bigdeli2026reformer}. A parallel LLM-routing literature selects which model to invoke under cost constraints using matrix-factorization, similarity-weighted, and classifier routers~\cite{routellm,hybrid_llm,best_route}; these transfer naturally to reformulator selection.

\noindent \textbf{Benchmarking query-variant selection.} Earlier efforts to capture formulation variance, such as \citep{uqv100}, released human-generated variations to evaluate robustness rather than selection. Recently, \citep{arabzadeh2026qpp} evaluates QPP as a variant selector for RAG pipelines, releasing variants alongside decision-based metrics. Other resources instead release reformulated queries or query pairs as the primary artifact, optimized toward a fixed objective: QueryGym standardizes LLM-based reformulation methods, prompts, and evaluation~\cite{querygym}; \textit{Matches Made in Heaven} provides supervised query pairs constructed to guarantee effectiveness gains over MS~MARCO~\cite{nguyen2016ms,matches-made-in-heaven}; and Refairmulate extends this to fairness-aware reformulation~\cite{refairmulate}. \sys{} differs by targeting the selection decision, releasing the action--outcome matrix selectors require: variants, ranked lists, retrieval scores, and oracle labels.
\begin{table}[t] 
\centering 
\vspace{-1em}
\caption{\sys{} benchmark. Each query is evaluated across 11 candidate systems, 5 backbone models, and 3 retrievers, yielding 165 query–candidate–backbone–retriever outcomes per query.}
\vspace{-1em}
\small
\label{tab:scale} 
\setlength{\tabcolsep}{3pt} 
\begin{tabular}{lrr} 
\toprule Dataset & Queries & Outcomes \\ 
\midrule TREC DL   & 147   & 24,255 \\ 
BEIR Core & 2,861 & 472,065 \\ 
BRIGHT    & 749   & 123,585 \\ 
\midrule 
Total     & 3,757 & 619,905 \\ 
\bottomrule 
\end{tabular} 
\vspace{-1em}
\end{table}

% ---- MAIN RESULTS: DL + BEIR + BRIGHT (combined) ----
\begin{table*}[t]
\centering
% \vspace{-2em}
\caption{Main results for Qwen2.5-7B-Instruct under sparse BM25 retrieval (nDCG@10) across TREC DL, BEIR, and BRIGHT. Best non-oracle per column \best{bold}, second-best \secondbest{underlined}. Oracle is an upper bound (excluded from ranking).}
% \vspace{-1em}
\label{tab:main-dl-beir-bright-qwen25-7b-bm25}
\resizebox{\textwidth}{!}{%
\begin{tabular}{l cccc c ccccccc c ccccccc c}
\toprule
& \multicolumn{4}{c}{TREC DL} & & \multicolumn{7}{c}{BEIR} & & \multicolumn{8}{c}{BRIGHT} \\
\cmidrule(lr){2-5}\cmidrule(lr){7-13}\cmidrule(lr){15-22}
Selector & DL19 & DL20 & DL-H & DL Avg & & SciFact & ArguAna & COVID & FiQA & DBPedia & News & BEIR Avg & & Bio & Earth & Econ & Psych & Robot & StackO & Sust & BR Avg \\
\midrule
Oracle (upper bound) & 0.772 & 0.750 & 0.475 & 0.666 & & 0.801 & 0.512 & 0.851 & 0.364 & 0.514 & 0.584 & 0.604 & & 0.504 & 0.583 & 0.334 & 0.428 & 0.261 & 0.348 & 0.336 & 0.399 \\
Original query & 0.506 & 0.480 & 0.285 & 0.424 & & 0.679 & 0.397 & 0.595 & 0.236 & 0.318 & 0.395 & 0.437 & & 0.182 & 0.279 & 0.164 & 0.134 & 0.109 & 0.163 & 0.161 & 0.170 \\
Best-single (fixed) & 0.687 & 0.632 & 0.357 & 0.559 & & 0.718 & \best{0.434} & 0.742 & 0.246 & 0.401 & 0.478 & 0.503 & & \best{0.417} & \best{0.500} & 0.242 & \best{0.363} & 0.170 & 0.249 & 0.263 & \best{0.315} \\
Random & 0.522 & 0.497 & 0.293 & 0.437 & & 0.698 & 0.406 & 0.708 & 0.230 & 0.353 & 0.459 & 0.476 & & 0.326 & 0.380 & 0.186 & 0.239 & 0.135 & 0.215 & 0.212 & 0.242 \\
\midrule
BERT & 0.685 & 0.609 & 0.332 & 0.542 & & 0.718 & 0.401 & 0.676 & 0.220 & 0.375 & 0.451 & 0.474 & & 0.377 & 0.456 & 0.217 & 0.275 & 0.147 & 0.240 & 0.242 & 0.279 \\
SW-Ranking & 0.622 & 0.590 & 0.328 & 0.513 & & 0.715 & 0.406 & 0.695 & 0.238 & 0.369 & 0.452 & 0.479 & & 0.405 & \secondbest{0.497} & 0.239 & 0.340 & 0.146 & 0.231 & \best{0.265} & 0.303 \\
MF & 0.593 & 0.575 & 0.327 & 0.498 & & 0.696 & 0.396 & 0.661 & 0.235 & 0.363 & 0.429 & 0.463 & & 0.308 & 0.396 & 0.192 & 0.222 & 0.132 & 0.225 & 0.201 & 0.239 \\
Pre-QPP (best) & 0.660 & 0.633 & 0.328 & 0.540 & & 0.718 & \secondbest{0.432} & 0.678 & 0.239 & 0.400 & 0.458 & 0.488 & & 0.399 & 0.499 & 0.242 & 0.327 & 0.162 & \secondbest{0.254} & \secondbest{0.263} & 0.307 \\
Post-QPP (best) & 0.679 & 0.610 & \secondbest{0.363} & 0.551 & & 0.712 & 0.427 & 0.659 & 0.237 & 0.394 & 0.454 & 0.481 & & 0.384 & 0.492 & \best{0.245} & 0.320 & 0.155 & 0.237 & 0.261 & 0.299 \\
Neural-QPP (best) & \best{0.717} & 0.632 & 0.362 & \secondbest{0.570} & & 0.715 & 0.423 & 0.734 & 0.248 & 0.389 & 0.461 & 0.495 & & 0.376 & 0.451 & 0.223 & 0.301 & 0.155 & 0.220 & 0.237 & 0.281 \\
LLM-as-judge & \secondbest{0.702} & \best{0.688} & \best{0.375} & \best{0.588} & & \best{0.723} & 0.417 & \best{0.762} & \best{0.275} & \best{0.424} & \best{0.489} & \best{0.515} & & \best{0.434} & 0.480 & 0.227 & \secondbest{0.361} & \best{0.189} & \best{0.257} & 0.248 & \secondbest{0.314} \\
\bottomrule
\end{tabular}%
}
% \vspace{-1em}
\end{table*}

% \vspace{-1em}
% # pending
\section{Query Reformulation Selection}

% \subsection{Task Formulation}

% Let $q$ denote an original user query, $\mathcal{C}$ a document collection, and $R$ a fixed retriever. A set of reformulation methods $\mathcal{E}=\{E_1,\ldots,E_M\}$ produces candidate query variants:
% \[
% \mathcal{Q}(q)=\{q_0,q_1,\ldots,q_M\},
% \]
% where $q_0=q$ is the original query and $q_i=E_i(q)$ for $i>0$. Each candidate induces a ranked list:
% \[
% L_i = R(q_i;\mathcal{C})=\langle d_{i,1}, d_{i,2},\ldots\rangle .
% \]
% A selector $S$ receives the original query, the candidate set, and any permitted features, and returns an index $\hat{i}=S(q,\mathcal{Q}(q))$. The selected ranking $L_{\hat{i}}$ is evaluated against qrels using a target metric $m$.

% For each query and target metric, we define the oracle selector:
% \[
% i_m^\star(q)=\arg\max_{i\in\{0,\ldots,M\}} m(L_i, qrels_q).
% \]
% The oracle defines the upper bound of the released candidate pool. The selector evaluation problem is therefore to approximate $i_m^\star(q)$ without access to test-set relevance labels.

We formulate query reformulation selection as a finite-action decision problem over a fixed pool of query variants. Let $q$ denote an original user query, $\mathcal{C}$ a document collection, and $R$ a fixed retriever. A set of reformulation methods
\[
\mathcal{E}=\{E_1,\ldots,E_M\}
\]
produces a candidate pool
\[
\mathcal{Q}(q)=\{q_0,q_1,\ldots,q_M\},
\]
where $q_0=q$ is the original query and $q_i=E_i(q)$ for $i>0$.
Each candidate variant induces a ranked list under the retriever:
\[
L_i = R(q_i;\mathcal{C})=\langle d_{i,1},d_{i,2},\ldots\rangle .
\]
For a target metric $m$, the true utility of candidate $q_i$ is
\[
u_i(q;R,m)=m(L_i,\mathrm{qrels}_q).
\]
The oracle-best candidate set is then
\[
I_m^\star(q;R)=
\left\{
i \in \{0,\ldots,M\} :
u_i(q;R,m)=\max_{j\in\{0,\ldots,M\}}u_j(q;R,m)
\right\}.
\]
When a single oracle label is needed, we use a deterministic tie-breaking rule, but the evaluation metadata retains all tied oracle-best candidates. A selector $S$ receives the original query, the candidate pool, and any features permitted by the benchmark protocol:
\[
\hat{i}
=
S\!\left(q,\mathcal{Q}(q),\Phi(q,\mathcal{Q}(q),R)\right),
\]
where $\Phi$ denotes optional features such as query statistics, retrieval scores, ranked-list properties, learned embeddings, or judge-derived evidence labels. The selected output is the ranked list $L_{\hat{i}}$. The goal of query reformulation selection is to approximate the oracle-best candidate without access to test-set relevance labels:
\[
\hat{i}(q) \approx i_m^\star(q;R).
\]

Selection is useful only when variants of the same query induce meaningfully different retrieval outcomes. For an instance $(q,\mathcal{C},R,m)$, the routeability of the candidate pool can be summarized by two gaps: the \emph{oracle headroom} over the original query,
\[
\Delta_0(q)=\max_i u_i(q;R,m)-u_0(q;R,m),
\]
and the \emph{fixed-policy gap} over the best fixed reformulator,
\[
\Delta_{\mathrm{bs}}(q)=\max_i u_i(q;R,m)-u_{\mathrm{bs}}(q;R,m).
\]
When $\Delta_{\mathrm{bs}}(q)$ is large, no fixed strategy matches per-query selection; when both gaps are small, a fixed choice already suffices.

\vspace{-0.5em}
\section{The Proposed \sys{} Benchmark}
\label{sec:benchmark}

% What the benchmark is for.
% Include Table~\ref{tab:benchmark_coverage}.

% \subsection{Purpose and Utility}
% \sys{} supports reproducible research on query variant selection. Each benchmark instance is an original query paired with a fixed pool of reformulated variants; a selector chooses one candidate, and the benchmark evaluates the ranked list it induces.  This setup isolates the selector as the object of study: competing methods operate over identical candidate pools, ranked lists, and evaluation labels.
% The resource supports three use cases. \textbf{Selector benchmarking}: methods are compared over identical candidate pools and ranked lists, so differences reflect the selector rather than the candidate generation or retrieval setup. \textbf{Supervised selector training}: per-query oracle labels provide the target for classifiers and routers. \textbf{Diagnostic analysis}: by comparing original-query, best-fixed-reformulator, and per-query oracle performance, researchers can inspect when selection is easy, hard, or unnecessary.

\subsection{Purpose and Utility}
\sys{} supports reproducible research on query reformulation selection by fixing the artifacts that are otherwise expensive and inconsistent to recreate---candidate reformulations, ranked lists, retrieval scores, oracle labels, and evaluation splits---so that competing methods are compared over the same action space and retrieval outcomes.

\vspace{-0.5em}

\subsection{Benchmark Construction}
\label{sec:benchmark-construction}

\sys{} materializes the action--outcome matrix used by selectors. For each query, retriever, and target metric, we store the candidate pool, the ranked list induced by each candidate, its retrieval utility, and the oracle-best candidate set. Formally, for each query $q$ we produce $(\mathcal{Q}(q), \{L_i\}, \{u_i\}, I^\star_m(q))$, where $\mathcal{Q}(q)$ is the candidate pool, $L_i$ is the ranked list produced by candidate $i$, $u_i$ is its utility, and $I^\star_m(q)$ is the set of oracle-best candidates under metric $m$. Construction proceeds in two stages.

\noindent\textbf{Variant Generation.}
For each query, we construct a pool $\mathcal{Q}(q)=\{q_0,\ldots,q_M\}$ containing the original query $q_0$ and $M$ LLM-generated reformulations. The reformulations span keyword-level methods (GenQR, GenQR-Ensemble, Query2E), document-level methods (Query2Doc ZS/FS/CoT, QA-Expand, MuGI), and corpus-grounded methods (CSQE, LameR), all generated through QueryGym~\cite{querygym} under a shared prompting and decoding interface. We generate these pools with five backbones---Qwen2.5-7B, Qwen2.5-72B~\cite{qwen7b}, Llama-3.1-8B, Llama-3.3-70B~\cite{llama3}, and GPT-4.1~\cite{openai2024gpt4technicalreport}---to support backbone-transfer analysis. Results for all models and retrieval settings are available in our repository.

\noindent\textbf{Retrieval.}
Each candidate is issued to three first-stage retrievers spanning lexical, learned-sparse, and dense retrieval: BM25, SPLADE, and BGE~\cite{splade, bge}. Retrieval runs through Pyserini~\cite{pyserini} over fixed indexes. For every query--candidate--retriever triple, we store the ranked list and scores, enabling selectors that use ranked-list or score-based features without rerunning retrieval.

\noindent\textbf{Datasets.} \sys{} spans three regimes of increasing difficulty. Web search uses TREC DL19/DL20/DL-Hard over MS~MARCO~\cite{TREC2019,TREC2020,DL_HARD,nguyen2016ms} with graded judgments. Zero-shot retrieval uses six BEIR datasets (SciFact, ArguAna, TREC-COVID, FiQA, DBPedia, TREC-News)~\cite{kamalloo}. Reasoning-intensive retrieval uses seven BRIGHT subsets~\cite{bright}, where relevance requires multi-step reasoning, so retrievers score lower and query reasoning helps disproportionately. Across these, five backbones, ten reformulators, and three retrievers yield the scale in Table~\ref{tab:scale}.

\vspace{-1.5em}
\subsection{Baseline Selectors}
We include four reference policies: \textsc{Original}, which uses the unreformulated query; \textsc{Best-single}, the dataset-level reformulator with the highest mean \ndcg{} for a given retriever; \textsc{Random}, which samples uniformly; and \textsc{Oracle}, the per-query upper bound. These baselines distinguish gains from reformulation itself from gains due to adaptive selection.

\noindent\textbf{Supervised classifiers.}
We treat query reformulation selection as a multiclass prediction problem, where each training query is labeled with its oracle-best reformulator under the target metric and retriever. We train a BERT-style classifier from the original query text from MSMARCO~\cite{nguyen2016ms} and evaluate it in a cross-collection setting.

\noindent\textbf{Similarity and matrix-factorization routers.}
We adapt model-routing methods to reformulator selection by treating each reformulator as an expert. The similarity-weighted router embeds queries with BGE~\cite{bge} and selects reformulators that performed well on nearby training queries. The matrix-factorization router learns latent query and reformulator representations from oracle-derived preferences. Both methods select from query-side signals and do not inspect retrieved documents at test time.

\noindent\textbf{Query performance prediction selectors.}
QPP selectors estimate the expected effectiveness of each candidate variant and choose the highest-scoring one. We include pre-retrieval predictors based on query and collection statistics, post-retrieval predictors based on ranked-list and score-distribution signals, and neural predictors based on BERT-QPP-style query--document interactions~\cite{kwok1996new,zhao2008effective,he2004inferring,ponte2017language,cronen2002predicting,zhou2007query,shtok2012predicting,tao2014query,arabzadeh2021bertqpp}. For compact reporting, \textsc{Pre-QPP (best)}, \textsc{Post-QPP (best)}, and \textsc{Neural-QPP (best)} denote the strongest predictor identified within each QPP family after evaluation; these rows are diagnostic within-family upper bounds rather than deployable selector choices.

\noindent\textbf{Outcome-level judge selectors.}
Outcome-level judge selectors estimate the utility of the ranked lists induced by each candidate. We use UMBRELA-style graded relevance labels in $\{0,1,2,3\}$ to score retrieved documents and select the candidate with the highest estimated ranking utility~\cite{upadhyay2024umbrela,farzi2025umbrela}.  Documents are judged against the original query rather than each reformulated variant, allowing shared documents across candidate lists to be judged once. Because this family relies on additional relevance estimation, we report it separately when discussing compute requirements. We adopt \texttt{DeepSeek-V3}~\cite{deepseekv3} as the default judge following~\cite{farzi2025umbrela}.

\begin{table}[t]
\centering
\vspace{-1em}
\caption{\textbf{Cross-retriever and cross-regime transfer} for Qwen2.5-7B-Instruct (nDCG@10, averaged over datasets within each regime). Best non-oracle per column \best{bold}, second-best \secondbest{underlined}.}
\vspace{-1em}
\label{tab:cross-retriever}
\setlength{\tabcolsep}{4.5pt}
\scalebox{0.82}{
\begin{tabular}{l ccc ccc}
\toprule
& \multicolumn{3}{c}{TREC DL (avg)} & \multicolumn{3}{c}{BEIR (avg)} \\
\cmidrule(lr){2-4}\cmidrule(lr){5-7}
Selector family & BM25 & BGE & SPLADE & BM25 & BGE & SPLADE\\
\midrule
Oracle (upper bound) & 0.666 & 0.689 & 0.722 & 0.604 & 0.671 & 0.587 \\
Original query & 0.423 & 0.586 & \best{0.612} & 0.437 & 0.569 & \best{0.525} \\
Best-single (fixed) & 0.559 & \best{0.600} & \best{0.612} & \secondbest{0.503} & \best{0.585} & \best{0.525} \\
\midrule
Supervised & 0.542 & \secondbest{0.572} & 0.483 & 0.473 & \secondbest{0.575} & \secondbest{0.280} \\
SW-Ranking & 0.514 & 0.376 & 0.517 & 0.479 & 0.565 & 0.220 \\
MF & 0.498 & 0.418 & 0.523 & 0.463 & 0.563 & 0.271 \\
Pre-QPP (best) & 0.540 & 0.017 & 0.531 & 0.488 & 0.572 & 0.240 \\
Post-QPP (best) & 0.551 & 0.350 & 0.517 & 0.480 & 0.577 & 0.257 \\
Neural-QPP (best) & \secondbest{0.570} & 0.269 & 0.525 & 0.495 & 0.574 & 0.249 \\
LLM-as-judge & \best{0.588} & 0.396 & \secondbest{0.535} & \best{0.515} & 0.571 & 0.247 \\
\bottomrule
\end{tabular}}
\vspace{-1.5em}
\end{table}

\vspace{-0.5em}

\subsection{Evaluation Protocol}

\label{sec:protocol}
We report two metric classes. \textbf{Retrieval effectiveness} uses \ndcg{} as primary, with \mapk{} and \rk{} as extended metrics. \textbf{Decision quality} uses three compact diagnostics: \textsc{Near-Or} (selected candidate within $\varepsilon=0.05$ \ndcg{} of the per-query oracle) and \textsc{Help}/\textsc{Hurt} (fraction better/worse than the original query). Additional diagnostics are provided in our repository.

\vspace{-0.5em}

% Across all three regimes, the candidate pool exhibits substantial oracle headroom. On TREC DL, \textsc{Original} obtains 0.424 \ndcg{}, \textsc{Best-single} improves to 0.559, and the per-query \textsc{Oracle} reaches 0.666. On BEIR, the corresponding scores are 0.437, 0.503, and 0.604; on BRIGHT, they are 0.170, 0.315, and 0.399. These gaps show that reformulation is valuable, but also that no fixed reformulator closes the per-query oracle gap. Among evaluated selectors, \textsc{LLM-as-judge} is the strongest non-oracle method on TREC DL and BEIR, reaching 0.588 and 0.515 respectively. On BRIGHT, \textsc{Best-single} remains marginally strongest at 0.315, with \textsc{LLM-as-judge} close behind at 0.314, while \textsc{Pre-QPP} and \textsc{SW-Ranking} recover part of the gain at 0.307 and 0.303. Thus, \sys{} is clearly routeable, but current selector families still recover only part of the available oracle improvement.

\vspace{-0.5em}
\section{Benchmarking the \sys dataset}
\label{sec:exp}

\begin{figure}[t]
  \centering
  \vspace{-2.5em}
  \includegraphics[width=1\columnwidth]{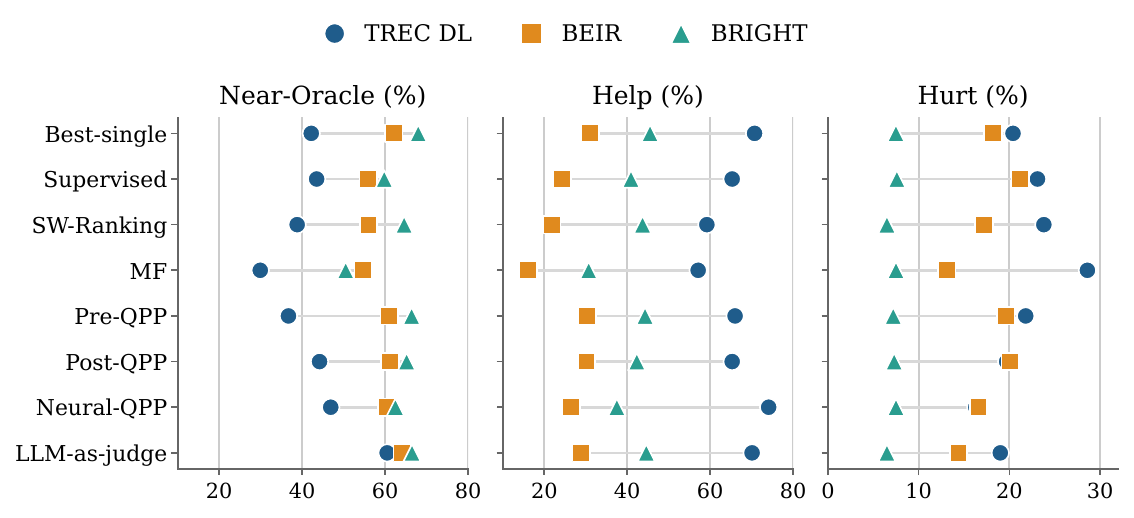}
    % \vspace{-2em}
  \caption{Decision quality on TREC DL/ BEIR/ BRIGHT. Averages are query-weighted within each regime. Hurt is lower better; higher is better for the other metrics.}
  \label{fig:decision-radar}
  \vspace{-2em}
\end{figure}

We use the released \sys{} artifacts to characterize how much headroom exists for query reformulation selection and how much of that headroom current selector families recover. Table~\ref{tab:main-dl-beir-bright-qwen25-7b-bm25} reports the main BM25 results using \primarybackbone. \textbf{First}, across all three regimes the pool shows substantial oracle headroom: on TREC DL, \textsc{Original}/\textsc{Best-single}/\textsc{Oracle} reach 0.424/0.559/0.666, with similar gaps on BEIR and BRIGHT. No fixed reformulator closes the oracle gap. \textsc{LLM-as-judge} is the strongest non-oracle selector on TREC DL (0.588) and BEIR (0.515); on BRIGHT it ties \textsc{Best-single} (0.314 vs.\ 0.315). \textbf{Second}, Table~\ref{tab:cross-retriever} shows that selector behavior is strongly conditioned on the retriever, and that the advantage of learned selection over fixed baselines does not transfer beyond sparse retrieval. Under BM25, \textsc{LLM-as-judge} is the best non-oracle method on both TREC DL (0.588 \ndcg{}) and BEIR (0.515), exceeding \textsc{Best-single}. Under BGE and SPLADE, however, no learned selector beats the fixed baselines: on TREC DL, \textsc{Best-single} reaches 0.600 under BGE while \textsc{Original} and \textsc{Best-single} tie at 0.612 under SPLADE; on BEIR, \textsc{Best-single} leads under BGE (0.585) and ties \textsc{Original} under SPLADE (0.525). The effect is most severe for BEIR/SPLADE, where learned selectors collapse to 0.22--0.28 against a 0.525 baseline. Selection therefore cannot be judged in a single sparse-retrieval setting: stronger retrievers can erase the benefit of adaptive reformulation. \textbf{Finally}, Figure~\ref{fig:decision-radar} shows that retrieval effectiveness alone does not fully characterize selector behavior. On TREC DL, \textsc{LLM-as-Judge} attains the highest \emph{near-oracle} rate of 60.5\%, but \textsc{Neural-QPP} provides a better help/hurt profile, with 74.1\% help and 16.3\% hurt compared with 70.1\% and 19.0\% for \textsc{LLM-as-Judge}. Similar trends appear on BEIR and BRIGHT, where selectors with comparable mean nDCG@10 exhibit different near-oracle, help, and hurt rates. For example, on BRIGHT, \textsc{LLM-as-Judge} and \textsc{Best-Single} achieve similar near-oracle rates (66.5\% vs. 68.0\%) with nearly identical hurt rates. These results show that mean effectiveness can mask query-level routing differences, motivating decision-quality diagnostics beyond aggregate retrieval metrics.

\vspace{-0.5em}
\section{Conclusion} 
\label{sec:conclusion} 

% We presented \sys{}, a resource for reproducible query reformulation selection that freezes the expensive action--outcome matrix across web search, zero-shot, and reasoning-intensive regimes, turning selector evaluation into a read-only decision problem. Benchmarking representative selector families shows substantial oracle headroom over both the original query and the best fixed reformulator, but current selectors recover only part of it. Selector-family rankings shift across retrievers and regimes, and mean effectiveness can hide different routing behavior, motivating decision-quality metrics in addition to retrieval effectiveness. \sys{} therefore provides a shared substrate for comparing future selectors while making clear that its oracle is bounded by the released candidate pool. Expanding reformulator diversity, retriever coverage, backbone coverage, and multilingual evaluation are important directions for future releases.

We presented \sys{}, a resource that freezes the action--outcome matrix for query reformulation selection across web search, zero-shot, and reasoning-intensive regimes, turning selector evaluation into a read-only decision problem. Benchmarking representative selectors reveals substantial oracle headroom over both the original query and the best fixed reformulator, yet current selectors recover only part of it: family rankings shift across retrievers and regimes, and mean effectiveness can mask different routing behavior, motivating decision-quality metrics alongside retrieval effectiveness. The oracle is bounded by the released candidate pool; expanding reformulator, retriever, backbone, and multilingual coverage are key directions for future releases.

% We introduced \sys{}, a resource that turns query reformulation selection into a cheap, reproducible benchmark by freezing the expensive action--outcome matrix---variants, ranked lists, retrieval scores, oracle labels, QPP features, and optional judge labels---across web search, zero-shot, and reasoning-intensive regimes. A new selector is evaluated by reading the released artifacts and emitting one candidate per query, with no regeneration, retrieval, or judging. Benchmarking four selector families under a shared decision-quality protocol shows how much selector-family ranking shifts across retrievers and regimes, and that mean effectiveness can mask substantially different routing behavior---findings that single-corpus or single-retriever studies cannot surface. By releasing the frozen substrate and harness, \sys{} provides common ground on which future selectors can be compared, at the same time, conclusions are necessarily conditioned on the released reformulator pool and artifacts: the oracle represents an upper bound over realizable choices within the action space rather than over all possible reformulations. Expanding reformulator diversity, retriever coverage, and cross-lingual evaluation remains an important direction for future releases.

\newpage
\section*{GenAI Usage Disclosure}

Large language models were used for light editing of author-written text, including grammar correction, improving fluency. All text reflects the authors' own ideas; no sections were produced entirely by a generative model. GenAI tools were also used to assist with coding tasks; all code was reviewed and validated by the authors

% \newpage
\balance

\bibliographystyle{ACM-Reference-Format}
\bibliography{sample-base}

\end{document}